\documentclass[superscriptaddress, prm, longbibliography, notitlepage]{revtex4-2}
\pdfoutput=1
\usepackage{graphicx}
\usepackage{bm}
\usepackage{natbib}
\usepackage{siunitx}
\usepackage{comment}
\usepackage{lineno} 
\usepackage[utf8]{inputenc}
\usepackage{cleveref} 
\crefname{equation}{Eq.}{Eqs.} 
\crefrangelabelformat{equation}{[#3#1#4--#5#2#6]}
\usepackage{gensymb}
\usepackage{xcolor}

\newcommand{\CFO}{CoFe$_2$O$_4$ }
\newcommand{\FO}{Fe$_3$O$_4$ }

\begin{document}
\title{Spin Canting in Exchange Coupled Bi-Magnetic Nanoparticles: Interfacial Effects and Hard / Soft Layer Ordering}

\date{26 April, 2021}

\author{C. Kons}
\affiliation{Department of Physics, University of South Florida, Tampa, Florida 33620, USA}

\author{K.L. Krycka}
\affiliation{National Institute of Standards and Technology, Gaithersburg, Maryland 20899, USA}

\author{J. Robles}
\affiliation{Department of Physics, University of South Florida, Tampa, Florida 33620, USA}

\author{Nikolaos Ntallis}
\affiliation{Department of Physics and Astronomy, Uppsala University, Uppsala, Sweden}

\author{Manuel Pereiro}
\affiliation{Department of Physics and Astronomy, Uppsala University, Uppsala, Sweden}

\author{Manh-Huong Phan}
\affiliation{Department of Physics, University of South Florida, Tampa, Florida 33620, USA}

\author{Hariharan Srikanth}
\affiliation{Department of Physics, University of South Florida, Tampa, Florida 33620, USA}

\author{J.A. Borchers}
\affiliation{National Institute of Standards and Technology, Gaithersburg, Maryland 20899, USA}

\author{D.A. Arena}
\email[Corresponding author: ]{darena@usf.edu}
\affiliation{Department of Physics, University of South Florida, Tampa, Florida 33620, USA}

\maketitle

\section{Abstract}
We investigate the spatial distribution of spin orientation in magnetic nanoparticles consisting of hard and soft magnetic layers.  The nanoparticles are synthesized in a core / shell spherical morphology where the magnetically hard, high anisotropy layer is CoFe$_2$O$_4$ (CFO) while the lower anisotropy material is Fe$_3$O$_4$ (FO). The nanoparticles have a mean diameter of $\sim$9.2 - 9.6 nm and are synthesized as two variants: a conventional hard / soft core / shell structure with a CFO core / FO shell (CFO@FO) and the inverted structure FO core / CFO shell (FO@CFO). High resolution electron microscopy confirms the coherent spinel structure across the core / shell boundary in both variants while magnetometry indicates the nanoparticles are superparamagnetic at 300 K and develop a considerable anisotropy at reduced temperatures.  Low temperature \textit{M vs. H} loops suggest a multi-step reversal process.  Temperature dependent small angle neutron scattering (SANS) with full polarization analysis reveals a strong perpendicular plane alignment of the spins near zero field, indicative of spin canting, but the perpendicular alignment quickly disappears upon application of a weak field and little spin ordering parallel to the field until the coercive field is reached. Above the coercive field of the sample, spins orient predominantly along the field direction.  At both zero field and near saturation, the parallel magnetic SANS peak coincides with the structural peak, indicating the magnetization is uniform throughout the nanoparticle volume, while near the coercive field the parallel scattering peak shifts to higher momentum transfer (Q), suggesting that the coherent scattering volume is smaller and likely originates in the softer Fe$_3$O$_4$ portion of the nanoparticle.

\section{Introduction}

Magnetic nanoparticles (NPs) are complex systems where unique properties can emerge that differ greatly compared to bulk counterparts such as enhanced magnetocrystalline anisotropy or Curie temperature dependent on size and shape of NPs  \cite{bulk_vs_np, NP_size_bulk, TC_size_effects}. Such properties can be further tailored in core/shell NP structures by careful selection of the constituent materials where interfacial coupling plays a strong role in the exchange interaction between layers \cite{Fauth_XAS}. Indeed the role of exchange coupling in composite structures containing both soft and hard magnetic phases has been studied extensively over the years as an avenue for tuning magnetic properties \cite{soft_hard_materials_1, soft_hard_materials_2, soft_hard_materials_3, NP_tuning}. Such a method pairs the high anistropy of a hard magnet with the high moment of the soft material for use in a variety of applications including biomedicine \cite{Raja_syn, hyperthermia_soft_hard}, data storage \cite{ferrites_electronics} and rare-earth free permanent magnets \cite{soft_hard_magnets}. 

The miniaturization of magnetic systems into the nano-region results in large surface area to volume ratios and, thusly, increases the fraction of surface spins compared to those in the bulk \cite{NP_size_effects}. As such, the sensitivity of the magnetic system on surface contributions becomes crucial and surface effects become driving forces in determining the overall magnetic properties \cite{NP_surface_effects}. Surface spin disorder as a result of symmetry breaking in the crystal structure at the NP surface and altered exchange interactions lead to canted spins and a reduction in magnetic moment of the NP \cite{NP_size_bulk}. In general, surface spins are more susceptible to spin canting than interior moments and such canting has been observed experimentally in many ferrites and iron oxides NPs \cite{OG_spin_canting, Kath_Fe2O3, COF_SANS, Oberdick, ferrite_canting, canting_aniso, Peddis}. The degree of canting is influenced by anistropy \cite{canting_fe_co_ferrite}, intraparticle effects due to Dzyaloshinskii-Moriya interactions and NP size \cite{canting_size} allowing for possible ways to tailor spin disorder in core/shell NPs through careful selection of constituent materials and synthesis parameters.

Spinel ferrites of the form MFe$_2$O$_4$ where M is a divalent transition metal are an attractive class of ferrimagnets offering both soft and hard phases as well as a common crystal structure that allow for a high quality crystal interface between core and shell layers \cite{mag_NPs}. In such materials the metallic ions are located at either octahedrally (B) coordinated sites or those with a tetrahedral (A-site) geometry and can adopt a normal, inverse or mixed spinel structure \cite{spinels}. The ferrimagnetic nature of this class of materials arises from the anti-parallel alignment of spins on the A and B sites. Of importance to this work is the inverse spinel structure where Fe$^{3+}$ cations are equally distributed at both A and B sites while the divalent M ions are found only at octahedral sites. In this work, \CFO (CFO) was selected due to its high anisotropy that should limit the degree of canting as spins will be more tightly bound to the crystal lattice compared to \FO (FO), which has a relatively high moment but much lower anistropy. A common crystal structure and negligible differences in lattice constant between the two materials (8.40 \si{\angstrom} for FO, 8.39 \si{\angstrom} for CFO \cite{Coey_mag}) enables synthesis of high quality core/shell NPs. 

Two core/shell NP variants were examined; a conventional one in which hard \CFO is in the core surrounded by a magnetically softer shell composed of \FO (core@shell, CFO@FO) and the inverted structure (FO@CFO) where the higher anisotropy material is now in the shell. Although both NPs are composed of the same materials there are a number of intrinsic factors that may influence spin arrangement and interfacial coupling. Curvature effects between the two NP variants will be different at the interface for CS structures compared to other geometries \cite{FO_CFO_curvature, FO_CFO_curvature_nanocomposite}; in the CFO@FO system \CFO core will have concave curvature at the core/shell boundary while \FO shell will see convex curvature that is reversed in the FO@CFO structure. Recent reports have demonstrated the effect of curvature on magnetization in \CFO / \FO bent heteorostructures that may be applicable to core/shell NPs as well \cite{FO_CFO_curvature}. The outermost layer may see differences in surface chemistry or uncompensated spins affecting surface contributions to overall magnetic properties. Such effects can lead to non-colinear spin arrangements or uniformly canted layers that inhibit desired magnetic properties as magnetic moments no longer lie parallel to the applied field. The combination of magnetically hard and soft layers can also lead to magnetic proximity effects as the anisotropy of the \CFO layer can delay the onset of superparamagnetism in the magnetite \cite{CFO_BFO, magnetic_proximity_AFM_FIM, magnetic_proximity_FO_MO, magnetic_proximity_EB}.

The structural and magnetic systems of each NP variant were explored using a variety of techniques including transmission electron microscopy (TEM), field and temperature dependent magnetometry, as well as fully spin-polarized small angle neutron scattering (SANS). The latter is a technique capable of providing both structural and magnetic details of the individual core/shell layers as well as interparticle spin correlations. The ensemble average of both the magnitude and direction of magnetic moments can be resolved allowing for differentiation between reduced moments and presence of spin canting. Field and temperature dependent magnetometry were collected and compared to Langevin generated magnetization curves as such methods can also provide evidence of canted spins in the NPs. Micromagnetic simulations were also performed to discern the mechanisms responsible for spin canting in each \CFO / \FO NP variant. 

\section{Samples and Experimental Methods}

\begin{figure}
 	\includegraphics[width = 17 cm]{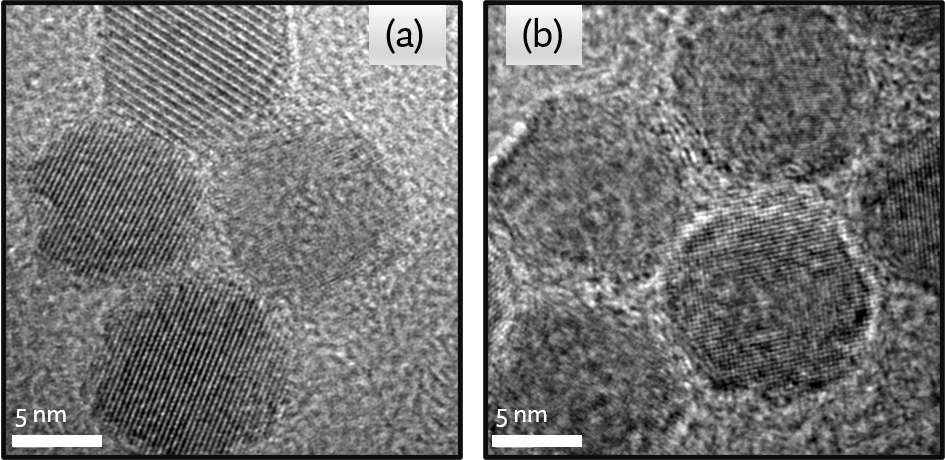}
 	\caption{\label{TEM} TEM images for the NPs used in the neutron scattering study of (a) CFO@FO and (b) FO@CFO core/shell NPs confirm the spherical shape and overall size. Owing to the thin shell and similar material densities between the core and shell layer it is not possible to see the individual NP layers with TEM. 
 	}
\end{figure}

Two variants of CoFe$_2$O$_4$ / Fe$_3$O$_4$ NPs were studied: using the notation of core@shell to represent the materials in each layer of the NP, one variant was composed of \CFO in the core and \FO in the shell layer (CFO@FO) while the inverted structure (FO@CFO) was the second NP studied. Both structures were synthesized via a similar seed mediated thermal decomposition process using 1,2 hexadecanediol, oleic acid (90\%), oleylamine (70\%), and benzyl ether (98\%) with iron (III) acetylacetonate to make \FO while adding cobalt (II) acetylacetonate to the mix would produce CoFe$_2$O$_4$ instead. For FO cores, the mixture was heated to 200$^\circ$C for 2 hours in a nitrogen gas environment before increasing temperature to 300$^\circ$C and refluxing for 1 hour. For the CFO cores, after the initial heating at 200$^\circ$C for 2 hours temperature is slowly increased at a ramping rate of 3$^\circ$C/min to 300$^\circ$C and refluxed for 30 minutes. The mixture is then cooled to room temperature where the cores are washed and collected by centrifuging with ethanol and hexane. 

The cores were then dispersed in hexane to be used as seeds for the shell layer growth; 85 mg of synthesized core were added to the above mixtures for Fe$_3$O$_4$ or CoFe$_2$O$_4$, depending on the desired shell composition. The core-mixture was then heated to 100$^\circ$C for 30 minutes to evaporate the hexane before refluxing at 300$^\circ$C for 1 hour. The now synthesized core/shell NPs were cooled to room temperature before being washed in ethanol and collected by centrifuging the mixture. The NPs were mixed with a small amount of hexane to prevent further oxidation or side reactions from occuring. This synthesis is known to produce NPs with a core 6 nm in diameter surrounded by a roughly 1 nm thick shell. Details on synthesis methods have been published elsewhere \cite{Joshua_synth, Joshua_synth_1, Joshua_synth_2}. A small portion of the powder was isolated for TEM studies of NPs size distributions and morphologies as well as magnetometry measurements. Figure \ref{TEM} shows each \CFO / \FO core-shell variant is primarily spherical but with faceting showing 6-fold symmetry. Due to the low Z-contrast between \CFO and \FO and shell layers close to the minimum resolution of the TEM it is not possible to see the individual core/shell layers. Lattice fringes are coherent across the core / shell structure indicating a coherent structure. 

Polarized SANS measurements were completed at NIST Center for Neutron Research using NG-7, a 30 m small angle neutron scattering instrument \cite{NG7}. A fully-polarized configuration was employed that uses an FeSi super mirror in front of the sample holder to initially polarize neutrons in a spin up orientation with an electromagnetic flipper coil that can reverse the spin direction of the incident neutron beam. A $^3$He cell located after the sample environment moderates the final neutron spin state by only allowing spins aligned with the $^3$He nuclear spins to pass; the direction of the $^3$He spins is reversed with a nuclear magnetic resonance pulse \cite{NIST_SS_2010}. In between both pieces of polarizing equipment is the sample environment where both \CFO / \FO NPs were mounted in an aluminum holder backfilled with He to prevent degradation of the samples during measurements. An electromagnet capable of horizontal fields up to $\sim$ 1.5 T was used for all measurements. On NG-7, the neutron beam is oriented in the Z direction and scattering in the XY plane is analyzed via a 2D detector (refer to Fig. 4 for coordinate system). The angular distributions of scattered neutrons was recorded by a detector whose position from the sample could be varied to a cover a range off scattering vectors ($Q$). The 2D scattering profiles were reduced using a custom Python script and analyzed in SASView 4.2.2 \cite{SASView_Joshua} using a custom core/multi-shell model. 

\section{Results and Analysis}
\subsection{Magnetic Studies}

\begin{figure}
	\begin{center}
 	\includegraphics[width = 17 cm]{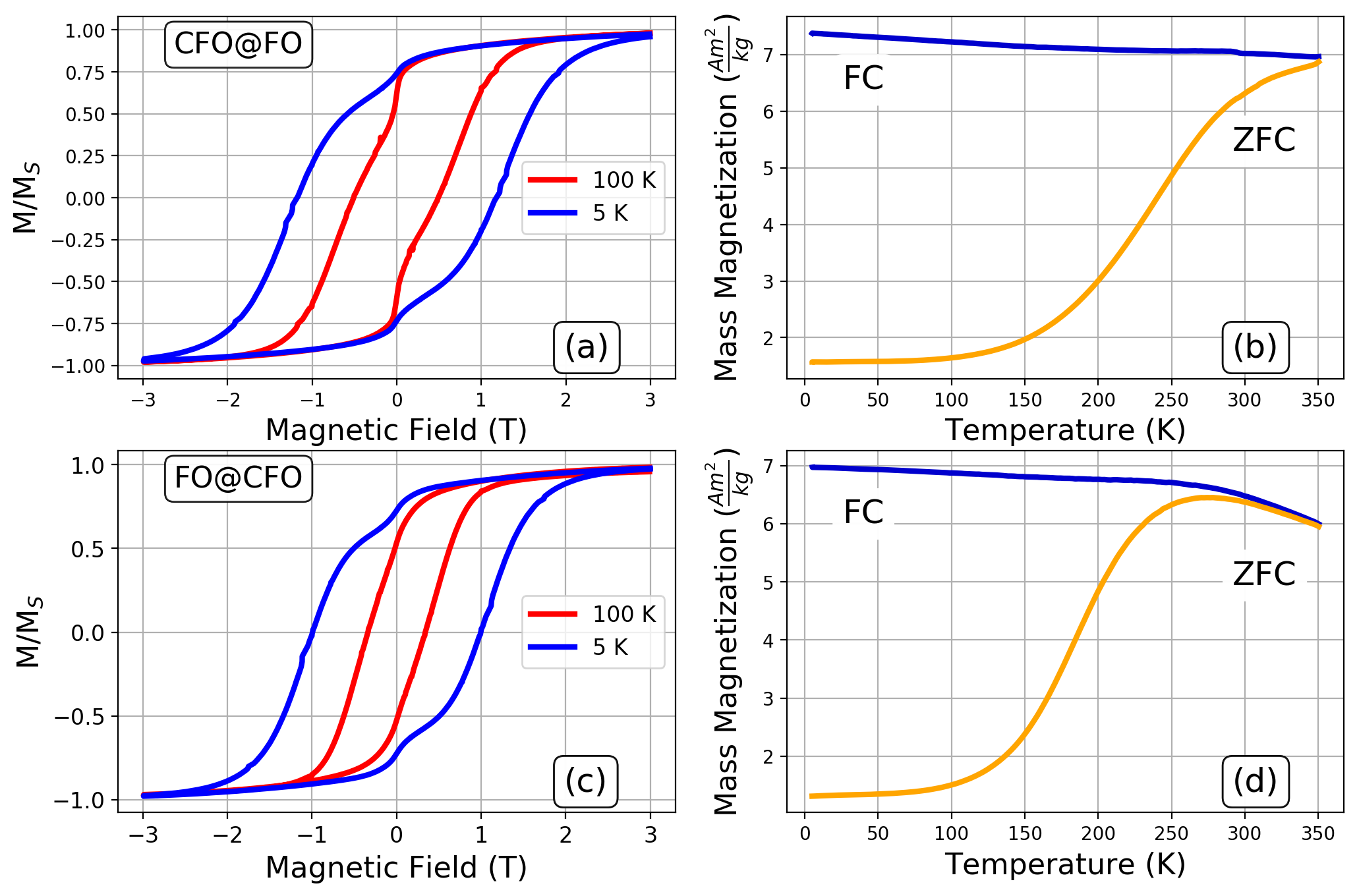}
 	\caption{\label{Magnetometry} Magnetization curves at 5 K and 100 K for (a) CFO@FO and (c) FO@CFO NPs with the corresponding field cooled (FC) and zero-field cooled (ZFC) M $vs.$ T curves show to the right for each NP. 
 	}
 	\end{center}
\end{figure}

Temperature and field-dependent magnetometry scans were collected for each NP ensemble; Fig. \ref{Magnetometry} shows hysteresis loops at 5 K and 100 K along with field-cooled (FC) and zero-field-cooled (ZFC) M $vs.$ T curves for CFO@FO and FO@CFO. Field cooling and other temperature-dependent magnetometry were performed in a 10 mT field for all measurements. From Fig. \ref{Magnetometry}(b,d) it can be seen that the blocking temperature for the CFO@FO NPs is well above room temperature, whereas, the FO@CFO NPs have a blocking temperature of $\approx$ 275 K. The hysteresis loops at 5 K and 100 K confirm the ferrimagnetic nature of both NPs below the blocking temperature with coercivity increasing as temperature decreases. 

Hysteresis loops at 5 K and 100 K in Fig. \ref{Magnetometry}(a) for CFO@FO NPs have similar saturation moments of 86 Am$^2$/kg and both show a secondary structure, or knee-like feature, between 0 and 1 T, suggesting a more complex spin reversal process between the core and shell. In the FO@CFO NPs this step is only present in the 5 K hysteresis loop as seen in Fig. \ref{Magnetometry}(c) while the 100 K loop shows no inflection at low fields; the saturation values are considerably different too, about 63 Am$^2$/kg at 100 K and increasing to 80 Am$^2$/kg at 5 K. The two NP variants also exhibit differences in coercivity indicating changes in exchange coupling based on the selection of hard or soft materials in the core and shell. At 5 K, when magnetically hard \CFO is in the core and paired with a softer \FO \ shell there is a coercive field $\sim$ 1.3 T (Fig. \ref{Magnetometry}(a)) while the inverted variant (\FO@\CFO) has a smaller coercive field of $\sim$ 1 T (Fig. \ref{Magnetometry}(c)).

\begin{figure}
	\begin{center}
 	\includegraphics[width = 17 cm]{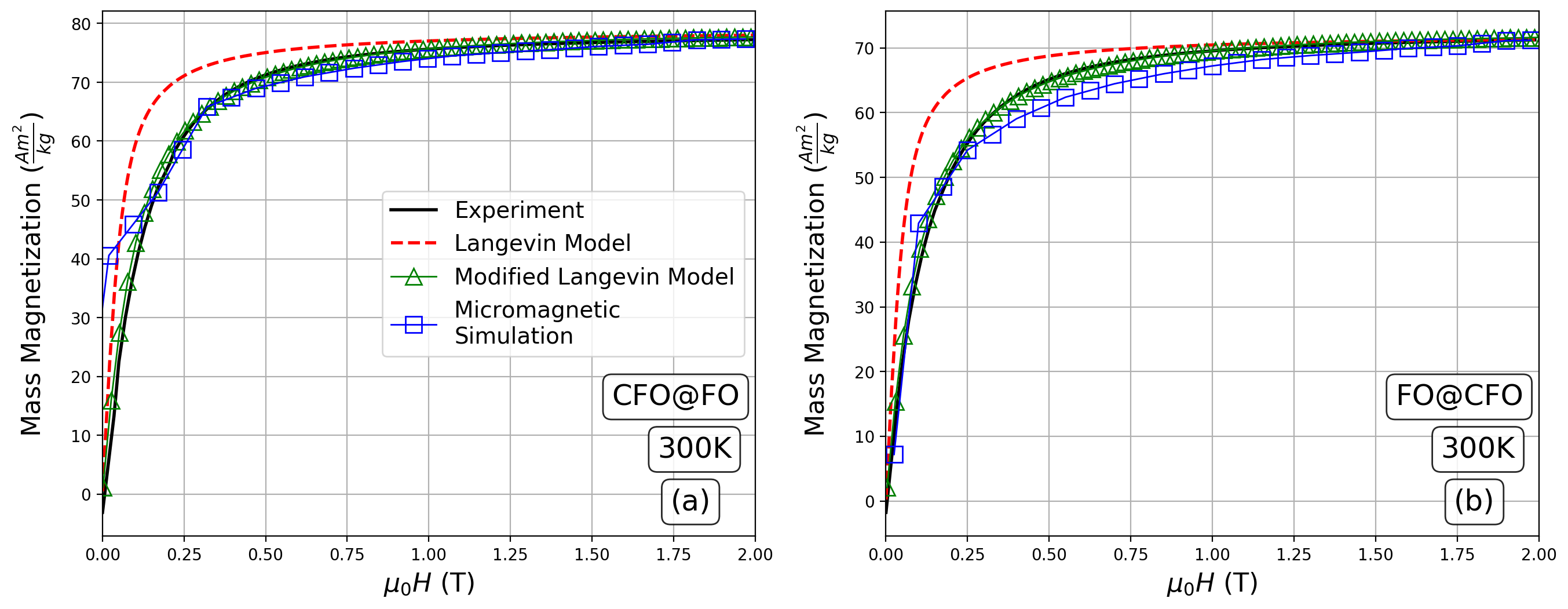}
 	\caption{\label{Langevin} Magnetization curves for (a) \CFO @ \FO  and (b) \FO @ \CFO at room temperature compared to the corresponding Langevin function and micromagnetic simulation.
 	}
 	\end{center}
\end{figure}

Reflecting the lower blocking temperature, the 
FO@CFO NPs are superparamagnetic (SPM) at room temperature while CFO@FO NPs have a small coercive field of $\approx$ 70 mT and remnant magnetization on the order of 3.5 Am$^2$/kg. Since both NP samples were in or very near the SPM state the magnetic response can be approximated using a Langevin function $\mathcal{L}(x)$ of the following form to simulate the M(H) curves:

\begin{equation}\label{MH_eqtn}
M(H)= n \, \mu \, \mathcal{L} \Bigg( \frac{\mu _0 H \mu}{k_B T} \Bigg) \qquad \textrm{where} \quad \mathcal{L}(x) = \frac{1}{\tanh (x) } - \frac{1}{x}
\end{equation}

where n is the total number of particles, $\mu$ is the magnetic moment per particle (Am$^2$), $\mu _0$H is the applied field (T), k$_B$ is the Boltzmann constant, and T is temperature \cite{Oberdick}. A value of $\mu$ was assigned for each NP based on structural parameters determined from SANS fittings and M$_S$ derived from experimental M(H) curves shown in Fig. \ref{Langevin}. In both NPs there are deviations between the experimental data and Langevin generated hysteresis loops below 0.8 T for CFO@FO  NPS and 0.7 T for FO@CFO variants. If uniform magnetization is assumed for each NP in Eq. \ref{MH_eqtn}  then the experimental values for $\mu$ would describe magnetic correlations at high and low fields. Near 0.1 T each NP sees $\approx$ a 25\% reduction in the experimental moment compared to the Langevin model indicating a reduction in the parallel magnetic component at low fields likely as a result of spin canting in a direction perpendicular to the field \cite{Oberdick}. 

The bare Langevin analysis assumes a uniform total moment per particle while TEM imaging and structural neutron scattering (below) indicate a variation of particle size and hence of moment per particle.  We also display in Fig. \ref{Langevin} a modified Langevin model that includes a log-normal distribution of nanoparticles (green triangles) \cite{Zohreh_NPs, Javi_langevin, Phan_Langevin}. The agreement with the measured $M$ $vs.$ $H$ curve is improved considerably, particularly in the low field region.  For the FO@CFO system (panel b), the magnetic diameter from the modified Langevin approach is 6.9$\pm$1.6 nm (1$\sigma$) while the magnetic diameter of the inverted CFO@FO variant comes in at 7.0$\pm$1.4 nm.  This is considerably smaller than the apprent size in the TEM images in Fig. \ref{TEM} and also smaller than the nanoparticle size estimates derived from the structural neutron scattering below.  One possible reason is that even at 300 K the particles are not completely in a superparamgnetic state, which is a fundamental assumption of both the Langevin and modified Langevin approaches.  Indeed, the CFO@FO nanoparticles exhibit a small coercivity at 300 K of nearly 6 mT.  Another possible contribution is spin canting in the near-surface region of the nanoparticles, which would tend to reduce the volume of the nearly single domain core region.  

To more closely capture the low field behavior of the nanoparticle assemblies and also introduce some degree of non-colinearity in the spin distribution across the nanoparticle, in Fig. \ref{Langevin} we also present calculations from micromagnetic simulations of the 300 K \textit{M vs. H} loops.  A simplified Hamiltonian is used for each nanoparticle that separates the core and shell into distinct macrospins and contains terms for the core / shell exchange and possible direct exchange between different nanoparticles (in the case of close contact between particles):

\begin{equation}
	\mathcal{H}=-\sum_{i,j}J_{ij}\bm{S}_i\bm{S}_j  -\sum_{i}k(cos)^2\bm{\theta}_i -g\mu_B\sum_i \bm{S}_i \bm{B} + \sum_{m,n}\bm{M}_m[\bm{D}]\bm{M}_n - \sum_{m,n}J_{inter}\bm{M}_m\bm{M}_n
\end{equation}

\noindent In this expression, $\{S_i\}$ represent the moment of macro spins within a nanoparticle while $\{M_n\}$ is the net moment of particle $n$, so that indices $\{i,j\}$ denote summation within a nanoparticle while $\{m,n\}$ denote interactions between nanoparticles. $J_{ij}$ is the Heisenberg interaction coupling between macrospins  $i$ and $j$ within a single nanoparticle. The magnetic anisotropy constant is represented by $k$ and  $\bm{\theta}_i$ is the angle between $\bm{S}_i$ and easy axis direction while $\bm{B}$ is the external magnetic field. The fourth term represents dipolar interactions between the net moment of each nanoparticle with moment $\bm{M}_n$ while $[D]$ is the dipolar tensor. The same notation holds for the last term of the Hamiltonian which describes interparticle interactions of the Heisenberg form with strength $J_{inter}$.  Temperature effects are simulated with the usual Boltzmann approach and Monte-Carlo methods are used to ensure sufficient sampling of the configuration space.  

For each nanoparticle, the total magnetic moment is subdivided into six macrospins, three each for the core and shell, and the sum of the macrospins for the core and shell equals the total moment per nanoparticle, as determined from bulk magnetometry.  The volume ratio of the core and the shell can be determined from the  structural (spin averaged) neutron scattering, as described below, and this ratio can be used to estimate the total moment per nanoparticle separately for the core and the shell.  For the CFO@FO variant, the core / shell moments are estimated as $4.14 \times 10^{-20}$ Am$^2$ / $12.38 \times 10^{-20}$ Am$^2$ [FO@CFO: $6.80 \times 10^{-20}$ Am$^2$ / $10.74 \times 10^{-20}$ Am$^2$].  The multiple macrospins per core and shell can simulate the effects of spin canting at the core / shell and shell / vacuum interface.  The ensemble of nanoparticles is simulated by placing the macrospins for each nanoparticle on a $12 \times 12 \times 12$ grid (\textit{i.e.} $12^3$ nanoparticles) with a mean spacing of 10 nm between nanoparticles and then imposing periodic boundary conditions.
 
Table~\ref{tab:Jij} and~\ref{tab:ki} present the parameters used for the macrospin model. For the different nanoparticles, indices $\{1-3\}$ refer to the core and  $\{4-6\}$ to the shell. In all cases, the anisotropy was assumed to be uniaxial for each particle, and its axis was randomly selected per particle in the assembly. $J_{inter}$ was set to $0.2$~mRy. To reduce the number of free parameters fit, one of the three  macrospins assigned per material, number $\{1\}$ for the core and number $\{4\}$ for the shell, is restricted to interact with the same exchange strength with the other two macrospins of the same material (see Table~\ref{tab:Jij} ). In this regime there are two exchange constants per material that have to tuned. In Table ~\ref{tab:ki} the anisotropy contributions are dissimilar between core and shell. The anisotropy of Fe$_3$O$_4$ shows a considerable enhancement when it is interfaced with CoFe$_2$O$_4$ thereby reducing the volume of the soft phase in each nanoparticle. This latter region of softer phase is the most probable nucleation region for the reversal process to start. Notably, according to this model the nucleation develops at the surface of the CFO@FO nanoparticles but in the core for the FO@CFO variation. In both cases the hard/soft interface acts as a strong pinning center which give rise to the large coercive field measured at low temperatures. 
 
\begin{table}[t]
 \caption{  Heisenberg interaction constant $J_{ij}$ for the macrospin model. All $J_{ij}$ are in units of $mRy$.  Vertical lines distinguish interactions within the core alone, between the core and shell, and solely in the shell.
 }
 \vspace{5pt}
 \centering
 \begin{ruledtabular}
 \begin{tabular}{lccc|cccc|ccc}
  \hline
  NP Variant &$J_{12}$&$J_{13}$&$J_{23}$&$J_{24}$&$J_{34}$ &$J_{25}$&$J_{35}$&$J_{45}$&$J_{46}$&$J_{56}$\\ 
  \hline
    
    CFO@FO & 3.4&3.4&-2.1&2.3&1.9&2.2&1.8&2.4&2.4&-1.8   \\
    FO@CFO  & 3.1&3.1&-1.8&2.1&2.2&2.3&1.7&2.8&2.8&-1.9   \\
    
 \end{tabular}
 \end{ruledtabular}
    \label{tab:Jij}
\end{table}

\begin{table}[t]
 \caption{  Anisotropy constant $k$ for the macrospin model. All $k$ are in units of $mRy$. $k_1$ through $k_3$ simulate the core while $k_4$ through $k_6$ correspond to the shell.
 }
 \vspace{5pt}
 \centering
 \begin{ruledtabular}
 \begin{tabular}{lccc|lccc}
  \hline
  NP Variant &$k_{1}$&$k_{2}$&$k_{3}$&$k_{4}$&$k_{5}$ &$k_{6}$\\ 
  \hline
    
    CFO@FO &  7.1 & 13.1 & 13.1 & 8.1 & 7.6 & 0.1   \\
    FO@CFO  & 0.8 & 6.4 & 6.4 & 9.2 & 8.1 & 8.0  \\
    
 \end{tabular}
 \end{ruledtabular}
    \label{tab:ki}
\end{table}

The micromagnetic calculations (blue squares in Fig. \ref{Langevin}) present better agreement with the magnetometry data than the bare Langevin approach, particularly for the FO@CFO variant where the superparamagnetic nature of the nanoparticles at room temperature is recovered and the approach to saturation is more gradual than the Langevin model.  The multiple macrospins per nanoparticle permit a simulation of spin canting that can arise in the nanoparticles, which is apparent in the magnetic SANS data below.  A similar slow approach to saturation is recovered for the CFO@FO variant, but the micromagnetic model suggests a weak coercivity (non-zero remanence) which is not observed in the data.   Additional details on the calculations, along with a fully atomistic calculation of the spins from a single nanoparticle, will be presented in forthcoming publication. 

%
\subsection{Small Angle Neutron Scattering}

\begin{figure}
	\begin{center}
 	\includegraphics[width = 17 cm]{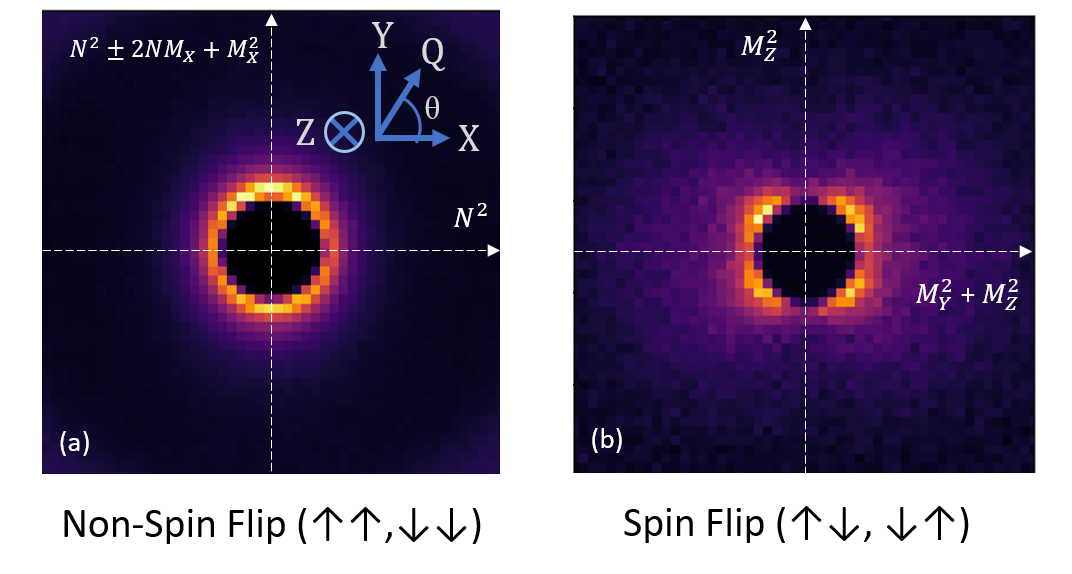}
 	\caption{\label{SANS 2D} Representative 2D scattering profiles for FO@CFO NPs in a 1.56 T field at 100K for (a) non-spin flip and (b) spin flip interactions showing nuclear and magnetic scattering contributions at key angles. Panel (a) also presents the coordinate system for these measurements.
 	}
 	\end{center}
\end{figure}

SANS measurements produce 2D scattering profiles where the intensity is related to the squared sum of the nuclear and magnetic Fourier transforms \cite{SANS_I_Fourier}. In a fully-polarized SANS setup the neutron direction is established before and after interacting with the sample allowing for measurements of all four scattering cross-sections; spin up ($\uparrow$) indicates a neutron that is aligned parallel to an external field while a spin down($\downarrow$) neutron is anti-parallel to the field. When a neutron maintains the same orientation before and after scattering from the sample it is known as a non-spin flip (NSF) interaction ($\uparrow \uparrow$ or $\downarrow\downarrow$) whereas a spin flip interaction measures cross sections where the neutron spin changes direction ($\uparrow \downarrow$ or $\downarrow\uparrow$). Representative 2D scattering profiles are shown in Fig. \ref{SANS 2D} for SF and NSF cross-sections along with scattering contributions at various angles. The 2D detector captures scattering in the X-Y plane orthogonal to the neutron beam, and sector cuts at key angles with widths of $\pm$ 10$\degree $ are averaged from these profiles. These sector cuts allow for calculation of individual nuclear (N$^2$) and perpendicular or parallel magnetic (M$^2 _{\perp}$ and M$^2 _{\parallel}$, respectively) scattering contribution where $\theta$ is the angle between the applied field in the X-direction and the scattering vector Q in the X-Y plane. The angular dependence of both nuclear and magnetic scattering results in the spin selection rules that simplify at key angles, allowing for analysis of individual scattering contributions. These angle-dependent polarization rules simplify as follows: 

\begin{equation}\label{SANS_N2}
N^2 (Q) = ( I^{\uparrow \uparrow}_{\theta = 0\degree} + I^{\downarrow\downarrow}_{\theta = 0\degree} ) 
\end{equation}

\begin{equation}\label{SANS_Mpara2}
M^2 _{\parallel} (Q) = \frac{( I^{\downarrow\downarrow}_{\theta = 90\degree} - I^{\uparrow \uparrow}_{\theta = 90\degree} )^2}{4 N^2}   
\end{equation}

\begin{equation}\label{SANS_Mperp2}
M^2 _{\perp} (Q) = \frac{1}{3} \left[ ( I^{\uparrow \downarrow}_{\theta = 0\degree} + I^{\downarrow\uparrow}_{\theta = 0\degree} ) + ( I^{\uparrow \downarrow}_{\theta = 90\degree} + I^{\downarrow\uparrow}_{\theta = 90\degree} ) \right]
\end{equation}
where ($\uparrow , \downarrow$) indicate scattering from spin up and down neutrons, respectively, and $I^{pq}$ corresponds to the scattering intensity along that sector cut angle for initial spin direction, $p$, and the selected spin direction after scattering, $q$, \cite{ NIST_SS_2010, Kath_Fe2O3, SANS_slxn_rules}. Each equation assumes isotropic nuclear scattering (N$^2$), M$^2 _{\parallel}$ = M$^2 _{X}$, and isotropic magnetic scattering orthogonal to the applied field M$^2 _{\perp}$ = M$^2 _{Y}$ = M$^2 _{Z}$ \cite{NIST_SS_2014}. Equation \ref{SANS_Mpara2} is sensitive to the net magnetization parallel to the applied field and M$^2 _{\parallel}$ will decrease at lower fields when spins are less ordered; the smaller magnitude of M$_\parallel ^2$ at low fields results in lower signal-to-noise ratios. 

Data were collected at temperatures ranging from 5-300 K and a variety of fields from 0-1.56 T to probe the magnetic configurations in the vicinity of the knee-like feature seen in the hysteresis loops (refer to Fig. \ref{Magnetometry} (a,c)). Each sample was cooled in zero field between temperatures since FC M $vs$ T curves showed negligible changes in magnetization. Samples were also magnetically trained to reduce random alignment of spins by first saturating the samples to positive saturation, reversing the field direction to negative saturation, back to zero field, and then measuring SANS patterns at the desired positive field value. 

\begin{figure}
	\begin{center}
 	\includegraphics[width = 17 cm]{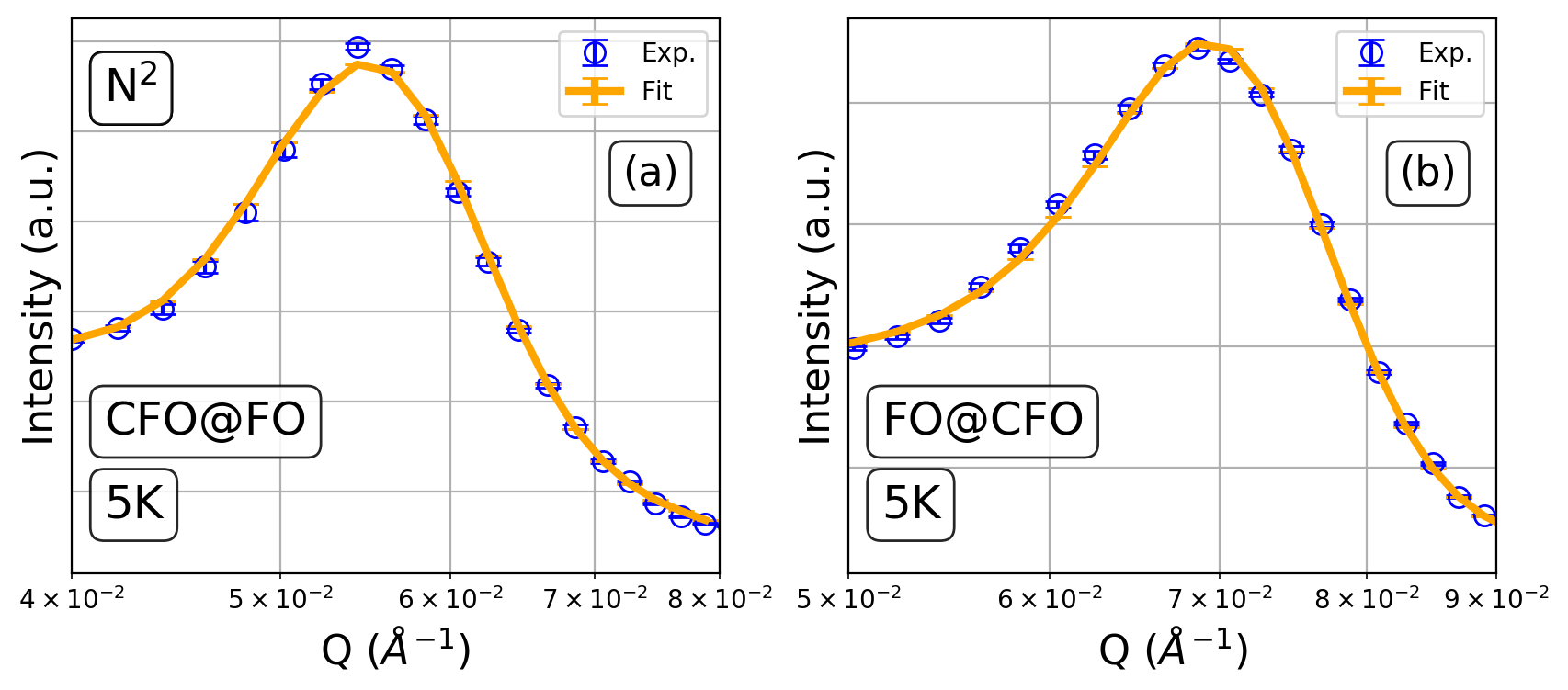}
 	\caption{\label{SANS N2} SANS nuclear (N$^2$) intensities for each CS NP variant as a function of the scattering vector Q at 5 K under HF (1.56 T) conditions. The solid lines show the best fit for each NP using a core + multi-shell model. Nuclear component were calculated as described in Eq. \ref{SANS_N2} using 10$\degree$ sector averages of the 2D data where $\theta$ is defined as the angle from Q in the  2D detector plane to the applied field oriented along the X axis.
 	}
 	\end{center}
\end{figure}

Structural parameters were determined by fitting the nuclear scattering; representative experimental data and fits at 5 K are shown in Fig. \ref{SANS N2} on a semi-log scale for each core/shell NP variant. Error bars representing 1-$\sigma$ distributions are shown in blue but are typically smaller than the marker size. As suggested by TEM micrographs the narrow size distribution of each NP type results in a well-defined Bragg peak that reflects the spacing of NPs. In CFO@FO NPs this Bragg peak is centered around  Q $\approx$ 0.055 $\mbox{\AA} ^{-1}$ for all temperatures and fields while scattering of the inverted variants reveal a peak near $\sim$ 0.07 $\mbox{\AA} ^{-1}$. The consistency in peak location and width indicates no structural changes for both NP samples with varying field or temperature conditions. A core + multi-shell model that assumed smooth, concentric layers devoid of surface roughness or interfacial mixing was used to fit the nuclear scattering data. The model assumes an outer layer composed of surfactants or other organic materials. Size distribution for each NP layer were also included in the model and shell layers exhibited negligible polydispersity values.

Modeling of CFO@FO core/shell NPs were completed over an extended Q-range in order to accurately determine background scattering contributions while Fig. \ref{SANS N2} focuses only on the region near the Bragg peak. Fitting results in an estimated average overall diameter of 10.4 $\pm$ 0.15 nm with a \CFO core radius of 2.9 $\pm$ 0.03 nm, \FO shell thickness of 1.7 $\pm$ 0.03 nm, and a 0.6 $\pm$ 0.02 nm thick surface layer. The inverted variant had a similar overall diameter of 10.2 $\pm$ 0.3 nm but different core and shell dimensions; the \FO core had a larger radius at 3.5 $\pm$ 0.05 nm and smaller subsequent shells than found in the CFO@FO NP; the CFO layer was calculated to be 1.3 $\pm$ 0.05 nm thick, and 0.3 $\pm$ 0.04 nm for the outermost layer. In CFO@FO NPs the average core/shell volume fraction is calculated to be $\frac{V_C}{V_S} = $ 0.33 while the inverted structure with a slightly larger core sees an increase in the core/shell volume fraction to about $\approx$ 0.63. 

Material compositions were confirmed with the model by comparing fitted SLD values to known values for \CFO and \FO. Fitted SLDs for \CFO in the core were found to be 5.9 $\pm$ 0.1 $\times 10^{-6} \mbox{\AA}^{-2}$ versus 5.5 $\pm$ 0.2 $\times 10^{-6} \mbox{\AA}^{-2}$ when in the shell, both are within 10\% of theoretical values \cite{NIST_SS_2014} and match well with reported values from similar NP systems \cite{Bonini_SANS, FO_CFO_SANS}. Similar variations in SLDs are seen for \FO in either the CFO@FO and FO@CFO NPs with $\rho$ = 6.8 $\pm$ 0.1 $\times 10^{-6} \mbox{\AA}^{-2}$ and 6.5 $\pm$ 0.1 $\times 10^{-6} \mbox{\AA}^{-2}$, respectively, deviating by less than 5\% with previously reported nuclear SLDs for Fe-oxides \cite{Fu}. The outer surface layer saw weakly negative SLDs ranging from -0.5 to -0.8 $\pm$ 0.2 $\times 10^{-6} \mbox{\AA}^{-2}$ in good agreement with known values for hexane \cite{Organic_SLDs}, supporting the assumption that this layer was formed from organic materials such as leftover surfactants from the NP cleaning process. The negative SLDs and thin outer layer indicate a relatively smooth surface with little roughness that would otherwise lead to higher nuclear SLDs due to interfacial mixing at the ferrite/surface boundary.

\begin{figure}
	\begin{center}
 	\includegraphics[width = 17 cm]{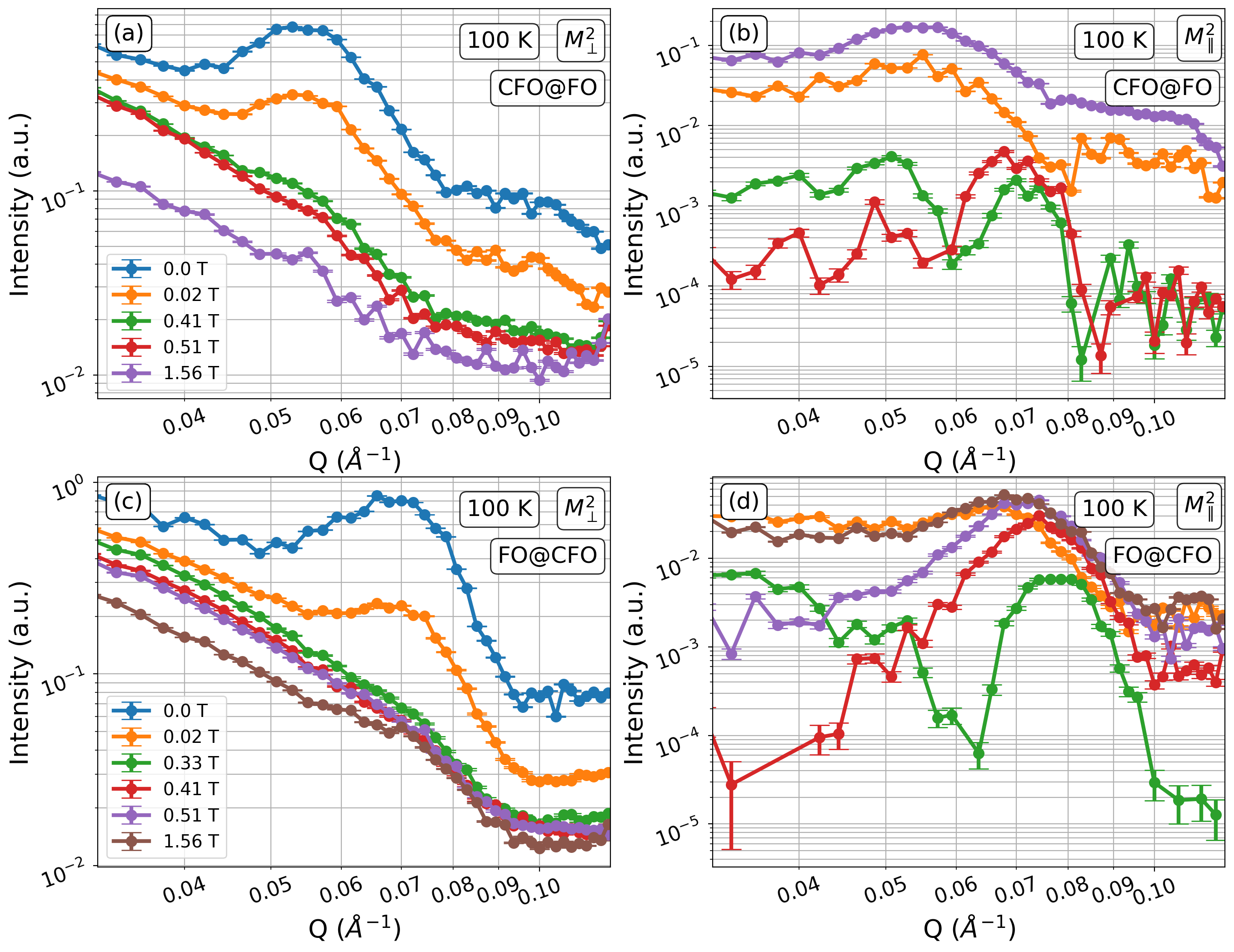}
 	\caption{\label{SANS Mag} SANS magnetic intensity data for each CS NP variant as a function of the scattering vector Q for 100 K and a variety of applied magnetic field conditions. Parallel (M$_\parallel ^2$) and perpendicular (M$_\perp ^2$) magnetic components relative to the applied field were calculated as described in \crefrange{SANS_N2}{SANS_Mperp2} using 10$\degree$ sector averages of the 2D data where $\theta$ is defined as the angle between the Q vector and the field oriented in the X direction.
 	}
 	\end{center}
\end{figure}

A fully-polarized SANS setup provides the capability to probe the 3D magnetic structure of NPs at variable fields allowing for determination of spin alignment \cite{SANS_spin_model}. Representative calculations of parallel (M$_\parallel ^2$) and perpendicular (M$_\perp ^2$) magnetic scattering cross sections from spin selection rules summarized in Eq. \ref{SANS_Mpara2} and \ref{SANS_Mperp2}; these data are presented in Fig. \ref{SANS Mag}. A similar story emerges for perpendicular spins (M$_\perp ^2$) of both NP variants (Fig. \ref{SANS Mag} (a,c)); at zero field a well defined magnetic Bragg peak is present, the peak intensity quickly diminishes at low fields and disappears entirely as the applied field increases. For both the CFO@FO and FO@CFO NPs the perpendicular magnetic Bragg peaks overlap with those from nuclear scattering indicating this spin alignment persists throughout the entire volume of the NP. 

\begin{figure}
	\begin{center}
 	\includegraphics[width = 17 cm]{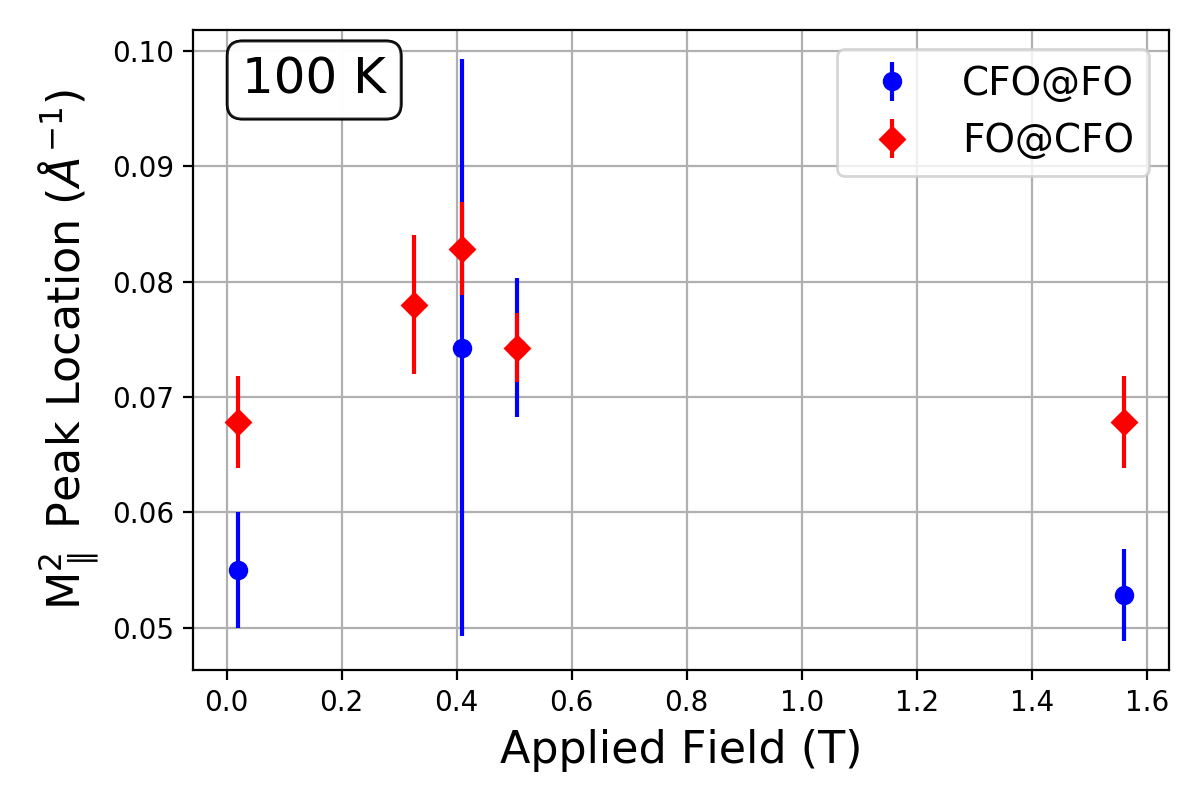}
 	\caption{\label{SANS Peak Loc} Peak location of M$_\parallel ^2$ cross section vs. field at 100 K for the two CoFe2O4 / Fe3O4 nanoparticle variant.  For both samples, there is a tendency for the peak to shift to higher Q near $H_c$ indicating a smaller magnetic volume.
 	}
 	\end{center}
\end{figure}

Fig. \ref{SANS Mag}(b) shows parallel spin ordering for CFO@FO NPs at various fields; at zero field no spin ordering in the direction of the field is observed (0 T data is below scale of graph), as the field is switched on a weak Bragg peak is observed (0.02 T). However, as the field is increased near H$_c$ (0.41 T) parallel magnetic ordering collapses, starts to reorder just above $H_C$ (0.51 T) at higher Q, and comes back strongly at high field (1.56 T). Parallel scattering in FO@CFO NPs, shown in Fig. \ref{SANS Mag}(d), indicates a more complex spin ordering process; similar to CFO@FO NPs no spin alignment is seen at 0 T with a Bragg peak not developing until low fields (0.02 T) but as the field is increased to near (0.33 T) and slightly above coercivity (0.41 T, and 0.51 T) the Bragg peaks shift to higher Q values until high field (1.56 T) where the peak returns to its original location. As mentioned previously, M$_\parallel ^2$ is representative of the net scattering between magnetic moments aligned both anti-parallel and parallel to the applied field. At low field values or near $H_C$ when spin alignment is weaker a higher signal-to-noise is to be expected resulting in Bragg peaks that are not as well-defined. This is illustrated in Fig. \ref{SANS Mag}(b) where the 0.41 T curve is indicative of some parallel spin ordering but the Bragg peak is distorted compared to higher field data. A summary of these peak locations with field is shown in Fig. \ref{SANS Peak Loc}.

\subsection{Discussion and Conclusions}

TEM micrographs of each \CFO / \FO NP variant shown in Fig. \ref{TEM} reveal nearly spherical NPs on the order of 10 nm in diameter. From Fig. \ref{SANS N2}, nuclear SANS scattering for each NP are well fitted using a using a spherical core + multi-shell model, corroborating the morphology determined from the TEM micrographs. Magnetometry of CFO@FO NPs [Fig. \ref{Magnetometry}(a,b)] indicate a blocking temperature well above room temperature while hysteresis loops at 5 and 100 K reveal $H_C$ increasing at lower temperatures. FO@CFO NPs, shown in Fig. \ref{Magnetometry} (c,d), have a lower blocking temperature around 275 K and smaller coercive fields values compared to CFO@FO variants. 

Neutron scattering analysis of each \CFO / \FO NP variant reveal the complex nature of magnetic ordering in these core/shell systems. The presence of parallel and perpendicular magnetic Bragg peaks in Fig. \ref{SANS Mag} for both NP systems is indicative of a coherent spin structure that persists across NPs instead of an isolated particle magnetic structure \cite{FO_MFO_spin_canting, Kath_Fe2O3, COF_SANS}. Referring to Fig. \ref{SANS Mag} and \ref{SANS Peak Loc}, at 0.33 T for FO@CFO NPs the parallel scattering Bragg peak shifts to higher Q with the peak location slowly decreasing back to lower Q values as the field increase; as scattering experiments probe samples in reciprocal space a tend towards higher Q value is correlated with a smaller scattering volume. At low and high fields the parallel magnetic Bragg peaks overlap with those arising from nuclear scattering indicating that the magnetic domains occur on the same length-scale as the overall NP structure but at intermediate fields, characterized by a M$^2 _{\parallel}$ peak at higher Q, spin ordering is confined to a smaller region of the NP. Correlating differences in magnetic scattering volume with the step-like features of the measured MH loops (Fig. \ref{Magnetometry}(a,c)) suggest that the spins in the core and shell do not become aligned in unison; spins in the magnetically soft magnetite core may align with an external field at a lower field than those in the harder shell layer. 

The knee-like features seen in the hysteresis loops of Fig. \ref{Magnetometry} have been reported in other hard/soft systems for both NPs and thin films alike as a result of exchange spring coupling across the hard/soft interface \cite{ soft_hard_materials_3, exchange_spring_hard_soft, NP_exchange_spring, NP_exchange_spring_2}. Exchange spring coupling is strongly tied to the dimensions of the soft phase; when the size of the soft region is less than twice the length of the domain wall of the hard phase ($2 \delta _H$) the system will see strong exchange coupling and a smooth hysteresis curve \cite{soft_hard_materials_1, soft_hard_materials_2, NP_hard_soft_exchange_magnet}. If the soft region is larger than this critical size the system will see coupling only at the core/shell interface and the soft layer will nucleate spin reversal at a lower field than the hard phase \cite{exchange_spring_hard_soft}. In core/shell NPs with overall dimensions below that of $2 \delta _H$ magnetic properties are still tied to the dimension of the soft phase; structures where the volume fraction of the soft material dominate have also been shown to exhibit two-phase hysteresis loops similar to those seen in Fig. \ref{Magnetometry} \cite{CFO_exchange_spring, NP_exchange_spring_2, NP_exchange_spring, exchange_spring_particles, weak_exchange_spring, NP_exchange_spring_effect}. In the CFO@FO NPs the softer magnetite shell comprised nearly 75\% of the overall NP volume and the magnetic effects are well explained by exchange spring coupling that pins spins near the core/shell interface while allowing spins near the surface to rotate freely as the external field is increased. In a strongly coupled system the spins in the core and shell regions behave coherently and so the spin state switching process will happen uniformly throughout the NP. Weaker coupling between core/shell layers can lead to pinning of shell spins near the interface while those closer to the surface are free to rotate independently of core and interface spins. 

With a blocking temperature between 40-50 K the magnetite layer will enter a SPM state long before the harder \CFO layer. In the FO@CFO variants the NP volume will be dominated by the shell layer ($\approx$ 61\% total volume); above T$_b$ of FO the large stray field of the CFO shell can lead to ordering of spins in the magnetite core while in the inverted CFO@FO structure CFO now occupies only 25\% of the total volume and the stray field may not be sufficient to influence ordering in the FO shell \cite{HafsaSpin, Joshua_synth}.

The spin evolution with field is more complex in the inverted FO@CFO system where the NP is no longer dominated by the soft phase as the magnetite core occupies only $\approx$ 39\% of the core/shell volume. From Fig. \ref{Magnetometry}(c) the knee-like features common in hysteresis loops of exchange-spring magnets are present at 5 K but noticeably absent at 100 K and, despite a greater hard phase volume fraction, the coercivities are lower than those seen for the CFO@FO NPs. In the conventional hard@soft variant the \CFO layer only has one interface for coupling to occur whereas in the inverted system when CFO is in the shell now interfacial effects can happen at the core/shell boundary and at the surface. As the Zeeman energy increases, spins in the FO core can easily switch direction while those in the CFO shell face competing spin disordering effects at both the core/shell interface and at the surface. A larger core also means more CFO spins will be located at the core/shell interface due to the increased surface area, in combination with disordering at two interfaces could result in a smaller measured coercive field. 

In summary, the spin distributions in core/shell NPs composed of \CFO and \FO were studied with fully-polarized SANS and conventional magnetometry. Magnetic scattering of both CFO@FO and FO@CFO NPs reveal multi-particle correlations in directions parallel and perpendicular to the field, spin canting is present at low fields but disappears at intermediate and higher applied fields. The shift of Bragg peaks with increasing field for parallel magnetic ordering strongly suggests the system behaves as an exchange-spring magnet where spins in the soft layer nucleate at lower fields compared to those coupled at the interface or in the harder \CFO layer. Evidence of a more complex spin reversal mechanism is reflected in the hystersis of each NP variant where a knee-like feature can be seen. Results indicate that in addition to properties such as material selection (which establishes the bulk magnetization and anisotropy of the constituents) and overall nanoparticle size, additional parameters including core / shell volume ratios and number of interfaces of the magnetically hard component (core to shell vs. shell to core + shell to surfactant layer) greatly affect the spin canting and overall spin ordering within the nanoparticles.  Control over these parameters will facilitate development of magnetic NPs with desired properties.

\section{Acknowledgments}

This material is based upon work supported by the National Science Foundation under Grant No. ECCS-1952957.  DAA acknowledges the support of the USF Nexus Initiative and the Swedish Fulbright Commission.  N.N. and M.P. acknowledge financial support from Knut and Alice Wallenberg Foundation through Grant No. 2018.0060. Some of the computations were performed on resources provided by the Swedish National Infrastructure for Computing (SNIC) at the National Supercomputer Center (NSC), Linköping University, the PDC Centre for High Performance Computing (PDC-HPC), KTH, and the High Performance Computing Center North (HPC2N), Umeå University.  

\section{References}
\bibliographystyle{apsrev4-2}
\bibliography{Manuscript}
\addcontentsline{toc}{section}{References}
\end{document}